**Chornobyl radiation spikes are not due to military vehicles disturbing soil**

Wood M.D.[1*], Beresford N.A.[1,2], Barnett C.L.[2], Burgess P.H.[3], Mobbs S.[4]

[1]*University of Salford, Salford, M5 4WT, United Kingdom*

[2]*UK Centre for Ecology & Hydrology, Lancaster Environment Centre, Bailrigg, Lancaster LA1 4AP, United Kingdom*

[3]*Radiation Metrology Ltd., 1A Highworth Rd., Faringdon, SN7 7EF, United Kingdom*

[4]*Eden Nuclear and Environment Ltd., Greenbank Road, Eden Business Park, Penrith, CA11 9FB, United Kingdom*

*Corresponding author: m.d.wood@salford.ac.uk

**Abstract**

On 25th February 2022, increased gamma radiation dose rates were reported within the Chornobyl Exclusion Zone (CEZ). This coincided with Russian military vehicles entering the Ukrainian part of the CEZ from neighbouring Belarus.  It was speculated that contaminated soil resuspension by vehicle movements or a leak from the Chornobyl Nuclear Power Plant (ChNPP) complex may explain these spikes in radiation dose rates.  The gamma dose rate monitoring network in the CEZ provides a crucial early warning system for releases of radioactivity to the environment and is part of the international safeguards for nuclear facilities.  With the potential for further military action in the CEZ and concerns over nuclear safety, it is essential that such anomalous readings are investigated. We evaluate the hypotheses suggested to explain the apparent gamma dose rate increases, demonstrating that neither military vehicle-induced soil resuspension nor a leak from the ChNPP are plausible. However, disruption of the Chornobyl base-station's reception of wireless signals from the CRMS network may potentially explain the dose rate increases recorded.

**Keywords**

Chernobyl; Gamma dose rate; Safeguards; detector response; military action; Russian invasion

1. Introduction

The 1986 accident at Chornobyl Reactor 4 remains the largest release of radioactivity to the environment in the history of nuclear power generation. In the weeks that followed, people were excluded from a 4700 km$^2$ area around the ChNPP that became known as the Chornobyl Exclusion Zone (Figure 1).  For nearly 36 years, human activity in the CEZ remained minimal and mainly confined to the central 'technical area' around the nuclear power plant complex.  Of the five other Chernobyl reactors, Units 1, 2 and 3 continued operating until 1996, 1991 and 2000 respectively (NEA, 2002); Units 5 and 6 were never completed. When the Russian military took control of the ChNPP complex and the 2600 km$^2$ of Ukrainian CEZ territory on 24th February 2022 (IAEA, 2022a,b), they became occupiers of an area with a substantial nuclear waste legacy. This legacy includes fuel from the decommissioned reactors, radioactive waste burial sites and an extensively contaminated surrounding environment.

At the time of the accident, Unit 4 contained approximately 7.4x10$^7$ TBq; approximately 15% of this activity was released into the environment (IAEA, 2007), the majority as short-lived radionuclides. The activity has reduced over time due to radioactive decay, but >5.2x10$^5$ TBq remains within the New Safe Confinement that now covers the reactor buildings (SNRIU, 2017). All fuel from Units 1-3 and spent fuel from the period of reactor operations, which started in 1977, is retained in interim storage facilities within the CEZ (SNRIU, 2017). Additionally, clean-up operations following the 1986 accident established approximately 800 radioactive material burial sites within the CEZ containing a total of 14,000 TBq (Smith and Beresford, 2005), some of which will now have decayed. The Chornobyl-derived radionuclides most prevalent within contemporary CEZ surface soils are $^{137}$Cs (Figure 1) and $^{90}$Sr, with lower activities of actinides (Am and Pu-isotopes) also present. The main radionuclide contributing to CEZ gamma dose rates is $^{137}$Cs, with activity concentrations in some CEZ soils likely the highest on earth; values in the range 10$^5$ to 10$^6$ Bq kg$^{-1}$ dry mass have recently been measured (Barnett et al., 2021; Beresford et al., 2020; Beresford et al., 2022).

*1.1 Gamma dose rate monitoring network in the CEZ*

The Comprehensive Radiation Monitoring and Early Warning System (CRMS) for Chornobyl includes a network of 67 gamma detectors, predominantly within the CEZ (Figure 1); one of these detectors appears to have been out of commission since May 2021. The CRMS detectors measure the gamma radiation dose rate (ionising radiation energy absorbed per unit time), which are reported in microsieverts per hour. These detectors are understood to have self-contained battery power units, enabling continuous gamma dose rate monitoring and wireless data transmission of readings to a base station in Chornobyl (Ukrainian Atom Instruments and Systems Corporation, 2022). Funded by the European Commission (European Commission, 2022) and operated by the State Specialised Enterprise 'Ecocenter', the CRMS replaced the earlier Automated Radiation Monitoring System (ARMS), which included 28 gamma detectors with cable communication that were retained as a backup system (Bondarenko, 2011). Results from the CRMS can normally be viewed online (e.g. from https://www.saveecobot.com/).

On 25$^{th}$ February the United Nations International Atomic Energy Agency (UN IAEA; IAEA, 2022a) issued a statement on military activity in the CEZ, noting a report from the State Nuclear Regulatory Inspectorate of Ukraine (SNRIU) that CRMS gamma dose rate measurements had increased up to 9.46 µSv h$^{-1}$. Reports in the press suggested that increases in gamma dose rate were up to twenty-fold above the normal baseline (Gill, 2022; Turner, 2022). SNRIU reported that Ecocenter experts connected the increases in gamma dose rate readings to heavy military vehicles resuspending contaminated soil (SNRU, 2022b). This was widely accepted as an explanation (e.g. SNRIU, 2022a; Gill, 2022; Sparks, 2022; Kim, 2022; World Nuclear Association, 2022; NEA, 2022), but without any attempt to validate it. There were also suggestions that a leak from the Chornobyl complex could have been the cause of the increase observed in gamma dose rates (Mousseau, 2022; Al Jazeera News Agencies, 2022; Polityuk and Crellin, 2022; Watts, 2022). In this paper we evaluate the suggested causes of the increases in gamma dose rate measurements recorded by the CRMS following the Russian invasion.

## 2. Methods

*2.1 Collating gamma dose rate data*

Data from the CRMS network can usually be accessed via an EcoCenter web portal, which is mirrored by other sites including saveecobot.com (CC BY 4.0). Although the EcoCenter web portal went offline c. 25$^{th}$/26$^{th}$ February, using data for 25$^{th}$ February we were able to confirm that https://www.saveecobot.com/ data accurately reflected data from the EcoCenter web portal. The https://www.saveecobot.com/ website provides individual detector data for a period of approximately 14 days, but these data have to be manually extracted. On 3rd March 2022, we extracted available CRMS network data from the https://www.saveecobot.com/ site, covering the period from approximately 17th February to 2nd March. With the exception of one detector, which appears to have been out of commission since May 2021 (i.e. at the time of the Russian invasion 66 of the detectors were operational), data were available for all monitoring locations.

*2.2 Modelling gamma detector response to soil disturbance*

To model the potential influence of soil disturbance on gamma dose rates measured by CRMS detectors, we assumed that each detector was mounted at 1 m above ground level (a standard mounting height for environmental monitoring; HMIP, 1995) and calculated the contribution of gamma photons to detector measurement up to a radius of 50 m. We assumed that, under normal baseline conditions, the dose rate is derived from gamma emitting radionuclides within the top 10 cm of the soil profile. From the limited data available on the distribution of radionuclides in soil profiles in the CEZ, we have assumed that the $^{137}$Cs activity concentration in the 0-5 cm layer is 14 times higher than that in the 5-10 cm layer (data from study of Jackson et al., 2004); this is based on data for 2002 so it is likely that our assumption gives a conservative estimate of the activity concentration in the 0-5 cm layer.

To provide a simple modelling system that we could use to explore the influence of soil resuspension on dose rate recorded by a detector, we approximated a relative photon contribution with distance from detector by calculating the circumference of 1 cm thick contaminated soil rings centred on the detector at 1 m intervals up to 50 m. For uniform activity per unit area of the ring, the available photon flux is proportional to the circumference. We then modelled the photon flux reaching the detector from each ring accounting for the decrease with increasing distance (inverse square law) and attenuation due to the path length in soil. We assumed an attenuation of 0.05 cm$^2$ g$^{-1}$ for the gamma energy of $^{137}$Cs (662 keV; Hubbell and Selzer, 2004) and a soil density of 1.6 g cm$^{-3}$. The 0.05 cm$^2$ g$^{-1}$ attenuation value is the mean of the mass attenuation coefficient, which models the probability of any photon interaction, and the mass energy-absorption coefficient, which describes the energy removed from a gamma beam. The mass attenuation coefficient ignores the contribution to dose rate of Compton scattered photons so overestimates the attenuation when air kerma rate is the measurand of interest. Conversely, the mass energy-absorption coefficient underestimates the effective attenuation because some of the Compton photons generated are backscattered. For 662 keV, the relevant interpolated values from NIST (Hubbell and Selzer, 2004) are 0.079 for the mass attenuation coefficient and 0.030 mass energy-absorption coefficient.

For each 1 m radius interval we modelled rings at 1 cm interval soil depths up to 10 cm depth, using the midpoint depth within each soil section (i.e. the 0 – 1 cm depth section was represented by a ring located at 0.5 cm depth); as depth increased, the path length in soil also increased resulting in greater attenuation of gamma photons. We also included attenuation in air taking the density of air to be 0.0012 g cm$^{-3}$. The relative total photon flux reaching the detector was estimated by summing the attenuated flux from each ring radius at each soil depth. This total relative photon flux was representative of the detector reading that would be expected under normal (i.e. undisturbed soil) conditions.

To evaluate the potential increase in dose rate reading due to disturbance of contaminated soil by military vehicles, we assumed a soil disturbance depth of 3 cm (Ayers, 1994). Highly conservatively we assumed that all of the top 3 cm of soil over a 50 m radius around the detector would be resuspended in air, whereas, in reality, the potential soil resuspension would be mainly limited to the track/wheel width of the vehicles. The number of gamma photons reaching the detector from the 3 – 7 cm depth of soil also increased as the attenuation by the upper 3 cm of soil was no longer present (effectively, the 3 –4 cm depth section became the top 0 – 1 cm under the assumed soil resuspension conditions).  We also modelled attenuation of gamma photons due to soil resuspended in air assuming that the top 3 cm soil layer was distributed throughout the air to a height of 2 m (the density of soil in air being 0.024 g cm$^{-3}$).  To determine the potential increase in gamma dose rate due to soil resuspension, we calculated the ratio of the total photon flux estimated under assumptions of resuspension to the total flux estimated for undisturbed conditions. These assumptions of soil resuspension are highly conservative.  For comparison to the gamma dose rate readings we are in effect assuming constant soil resuspension over a one-hour integration period.  Also, the resulting air mass soil load is unrealistically high compared to those that would arise even within a dust storm (Zhang et al., 2005). Because the soil loading in air was unrealistically high, we repeated the calculations assuming no attenuation of photons by resuspended soil (i.e. a worst case scenario). The ratio of the estimated total relative photon flux with the top three centimetres of soil resuspended in air to that for undisturbed conditions was calculated as a measure of the increase in dose rate due to resuspended soil.

These calculations assumed that the CRMS detector was mounted over soil, we repeated all calculations assuming the detector was mounted on a concrete plinth and that the soil under the concrete did not contribute to the dose rate reading recorded by the detector. To model this, we conservatively assumed a concrete base extending to 2 m radius around the detector (i.e. assuming no gamma flux from the ground over the first 2 m from the detector).

It is possible that some detectors are mounted higher than 1 m, such as on cabins used to house other air monitoring equipment.  The field of view of the detector (i.e. the area over which contamination in the soil will influence the detector reading) will increase if detectors are mounted higher.  However, for uniform contamination with distance from the detector, increasing the height of the detector would have negligible effect on the dose rate.  This is because there is only a small increase in path length through soil as the detector height above ground increases and there is minimal attenuation of gamma photons in air.

### 3. Results and discussion

*3.1 Gamma dose rate measurements*

We have determined that spikes in dose rates were recorded by 39 of the 66 operational CRMS detectors on 24$^{th}$ and/or 25$^{th}$ February (coinciding with the Russian invasion); three of these are located to the south, outside of the CEZ (Figure 2).  All detectors subsequently went offline, although not all at the same time and for different time periods.  After the spike dose rate, 30 of the detectors went offline immediately and the remaining nine rapidly returned to baseline dose rate readings (within 30 min to 3 h after the spike) before also going offline.  A detector in the CEZ that is part of a different (Ukraine-wide) monitoring network operated by the Ukrainian Hydrometeorological Institute (UHMI) also went offline during this period.  It appears that other radiation monitoring networks in Ukraine were offline during the same period for different lengths of time.  Some of the CEZ detectors came back online from 28$^{th}$ February, but by 3$^{rd}$ March the entire CEZ network was offline once again.

When detectors were operating on both the 24$^{th}$ and 25$^{th}$ February, the spike in dose rate recorded on the latter date was consistently higher. Given that the other 27 detectors went offline during this period, it may be that their gamma dose readings also peaked but were not reported. For the 15 detectors operating during the day on the 25$^{th}$ February, the peaks were all reported between 09:20 and 10:50. Dose rates recorded by the 55 detectors operating 28$^{th}$ February – 2$^{nd}$

March returned to baseline values as recorded before 24$^{th}$ February (Figures 2 and 3). From 3$^{rd}$ March 2022 until the time of writing (6$^{th}$ April 2022) the CRMS network has been offline.

The monitoring data show order of magnitude higher dose rates for some locations than were initially reported by SNRIU. The gamma detector at Ladyzychi, located approximately 30 km south east of the Chornobyl Nuclear Power Plant on the opposite side (east) of the Pripyat River, reported the greatest increase in dose rate (576 times higher than the baseline); this equated to a dose rate increase of 60 µSv h$^{-1}$ (Figures 2 and 4). The other detectors reporting highest additional dose rates (62 – 90 µSv h$^{-1}$) were all within the boundaries of the Chornobyl Nuclear Power Plant complex. However, in contrast to Ladyzychi, the ratio of peak dose rate to baseline dose rate for these detectors was only 12-38 times the baseline. Seven other detectors had peak to baseline dose rate ratios in the range 40 – 278. These were located throughout the CEZ (Figure 4). The detector reporting a dose rate of 278 times the normal baseline dose rate was located in the north west of the CEZ, close to the Belarusian border and about 30 km from the Chornobyl Nuclear Power Plant.

*3.2 Soil resuspended by vehicles caused the rise in gamma dose rate recordings?*

To determine whether increases in gamma dose rates above baseline could be explained by military vehicle movements disturbing contaminated soils, we modelled the potential increase in dose rate that could occur if the top 3 cm of contaminated soil was moved into the air mass as described above. Based on historical weather data available for Chernobyl town (https://www.timeanddate.com/weather/ukraine/chernobyl/historic), there had been no precipitation since 20$^{th}$ February 2022. Given the majority of the CEZ is sandy soil, it is reasonable to assume that the soil surface would be relatively dry throughout the 24$^{th}$ and 25$^{th}$ February, increasing the likelihood of some soil resuspension by vehicle movements. However, much of the CEZ is forested so the majority of the vehicle movements were likely along the asphalt roads or compacted unpaved roads.

For detectors mounted over soil and making highly conservative assumptions (e.g. the entirety of the top 3 cm of contaminated soil over a 50 m radius being suspended in the 2 m of air mass above the soil surface for a 1-hour period and no attenuation of gamma photons by resuspended soil) we estimated a maximum gamma dose rate increase of 2.2 times the baseline. For detectors mounted over a concrete base extending up to 2 m from the detector, the maximum estimated gamma dose rate increase was 3 times the baseline. If we include photon attenuation by resuspended soil particles in the air mass then we do not estimate an increase in gamma dose rate. Our highly conservative modelling approach demonstrates that the observed increases in gamma dose rates above baseline measurements across many of the detectors cannot be explained by contaminated soil resuspension in air due to military vehicle movement (Figure 4). Furthermore, the fifty-five detectors that came back online all reported gamma dose rates that had returned to normal baseline levels at a time when there would still have been considerable military traffic through the CEZ as it appears to have been used as a route to transport large numbers of troops and military equipment into northern Ukraine. We attempted to further evaluate spatial patterns of military vehicle movement using Sentinel-1 and Sentinel-2 satellite data, but the resolution was too poor to do this.

*3.3 Could a leak from the Chornobyl complex explain the increased gamma dose rates?*

Another potential explanation for increases in dose rates is a breach of contaminated material containment within the reactor complex or one of the radioactive waste storage facilities (Mousseau, 2022). If that was the case, it would be anticipated that dose rates would increase local to the source of release and spread over the surrounding area based on wind speed and direction.

There are weather data for the town of Chornobyl available for 24th February and the morning of 25th (https://www.timeanddate.com/weather/ukraine/chernobyl/historic). On 24th and 25th February, when detector readings peaked, wind speeds were low (0 - 6.4 km h$^{-1}$) and in a northerly direction. On 24th February, the highest dose rate increases (approximately 50 - 60 µSv h$^{-1}$) were for three detectors within the boundary of the Chornobyl Nuclear Power Plant (peak to baseline dose rate ratios were 7.3 to 28.7) (Figures 2 and 4). However, the first detector to show a peak response was approximately 17 km to the east of the Chornobyl Nuclear Power Plant at Chapaievka. This detector peaked at 12:30 to approximately 37 times the baseline reading (an increase of 3.2 µSv h$^{-1}$), then went offline before recording a dose rate of 7.4 µSv h$^{-1}$ on 25th February at 09:20 (an increase of 7.3 µSv h$^{-1}$ over baseline) (Figures 2 and 4). All the other highest peak dose rate to baseline measurements on 24th and 25th February were to the west of the Chornobyl Nuclear Power Plant, the closest being approximately 16 km away (Figure 4). There were also two detectors within the Nuclear Power Plant boundary which showed relatively small increases in dose rate (2.4 µSv h$^{-1}$ and 7.8 µSv h$^{-1}$). In summary the spatial pattern of the changes in the dose rate readings across the CEZ (Figures 2 and 4) does not support a release of radioactivity from the Chornobyl complex. The return of detectors to baseline levels from 28th February 2022 (Figure 3) also suggest no significant additional deposition of radioactivity.

*3.4 Electromagnetic radiation and wireless interference*

Electromagnetic radiation (or radiofrequency interference) from military activities (e.g. radar, electronic warfare) may have impacted on the CRMS detectors directly (Matisoff, 1990; Brown, 2022). However, the spatial pattern of gamma dose rate increases observed within the CEZ does not seem support this explanation; some detectors increased whereas neighbouring detectors did not (Figures 2 and 4). The highest increase (approaching 600-fold) was at Ladyzychi, along a minor road that only provides access to this and one other location. Therefore, there is no logical explanation as to why there would be significant military activity at this site. Also, if the changes were due to military-related electromagnetic interference directly affecting the detectors, anomalous readings would be expected across other detector networks in Ukraine where military activity has occurred. However, this was not observable in detector dose rates reported for those other networks on the https://www.saveecobot.com/ website.

A more plausible hypothesis may be that military activity affected the reception of detector wireless signals by the Chornobyl CRMS base-station (Hessar and Roy, 2016) and, if that were the case, we may not expect to see a spatial or temporal pattern in anomalous dose rate readings. For 28 locations (27 locations throughout the CEZ and one in Slavutych) it is understood that the cable-connected detectors from the earlier ARMS system continue to provide a backup system for the CRMS (Bondarenko, 2011). If the wireless CRMS network was affected on 24th and 25th February, dose rates reported on these dates from locations that were part of the earlier ARMS network may have come from the cable-connected detectors and hence would not be affected by issues related to wireless data transmission. At the time of the Russian invasion, 27 locations did not report peak dose rates and one detector location had been offline since May 2021 (Figure 5). Three of the detectors that did not report peak gamma dose rates went offline on 23rd February between 13:00 and 19:00. The remaining 24 were reporting until between 21:00 on 24th February and 01:00 on 25th February. This covered the period when some of the peak dose rates were reported by other detectors, but all of the detectors that did not report a peak were offline when the highest peaks were observed (09:20-10:50 on 25th February). Therefore, we cannot be sure whether or not they would have reported peak dose rates if they had been operational. It is not currently possible to verify which locations throughout the CEZ retain cable-connected backup detectors. However, the lack of dose rate change at Slavutych, which is known to have a cable-connected detector from the earlier ARMS network, may lend further credence to the hypothesis that military action affected the reception of detector wireless signals by the Chornobyl CRMS base-station.

## 4. Conclusions

Given the potential implications for human and environmental exposure to radiation, it is essential that deviations in dose rates recorded by the CRMS network can be adequately explained. Whilst military vehicle movements will undoubtedly have increased the dust loading in the air mass (Wellings et al., 2019), our analyses demonstrate that, contrary to wide speculation within the media and scientific community (Mousseau, 2022; Mothersill, 2022), such resuspension of contaminated soil cannot explain the gamma dose rate increases reported for many detectors in the CRMS network on 24$^{th}$/25$^{th}$ February 2022. The elevated readings also do not show a spatial pattern that may suggest they are a consequence of radioactivity releases from the Chornobyl nuclear power plant complex. Military electro-magnetic frequency interference may potentially cause reporting anomalies from detectors, but again this would be expected to follow a spatial pattern and to be observed for gamma dose rate detectors at other locations in Ukraine where military activity has taken place. A more plausible explanation may be that reception of wireless signals by the CRMS network base-station in Chornobyl was disrupted.

The current CRMS network and the earlier ARMS detectors have been providing a reliable source of gamma dose rate monitoring throughout the CEZ for over three decades (Bondarenko, 2011). The network provides a crucial early warning system for releases of radioactivity to the environment and is part of the international safeguards for nuclear facilities (IAEA, 2011; 2014). Anomalous readings were confined to the period of military activity (24$^{th}$/25$^{th}$ February), with dose rates returning to baseline levels after 28$^{th}$ February 2022. As stated by the International Atomic Energy Agency (IAEA, 2022c), military activity in the CEZ is of concern due to the quantity of radioactivity remaining in the area. Similar concerns remain for other operational reactor sites in Ukraine, such as the Zaporizhzhya Nuclear Power Plant (IAEA, 2022d). Wildfires, which are a common occurrence in the CEZ (Beresford, et al., 2021) and result in smoke contaminated with radionuclides, lead to concerns of increased exposure rates by local and wider European populations (Beresford, et al., 2021). Although in reality the risk is low (Beresford, et al., 2021) the CRMS network plays an important role in ensuring increased gamma dose rates from wildfire events can be evaluated; wildfires in the CEZ have been reported by the Ukrainian authorities in mid-March (SNRIU, 2022c) and others will occur throughout the year. At the time of writing, the CRMS network results could not be viewed on https://www.saveecobot.com/; the same applies to detector networks around other Ukrainian nuclear sites. The CRMS, and other monitoring networks, need to come back online urgently so that the radiological situation in Ukraine, where intense military activity continues in some areas, can be monitored.

**Acknowledgements**

We thank Dan Morton (UKCEH) for reviewing the Sentinel satellite data. We also thank our Ukrainian colleagues and friends, who have continued to provide updates on the situation in Chornobyl during this challenging period, we hope that they and their families continue to remain safe.

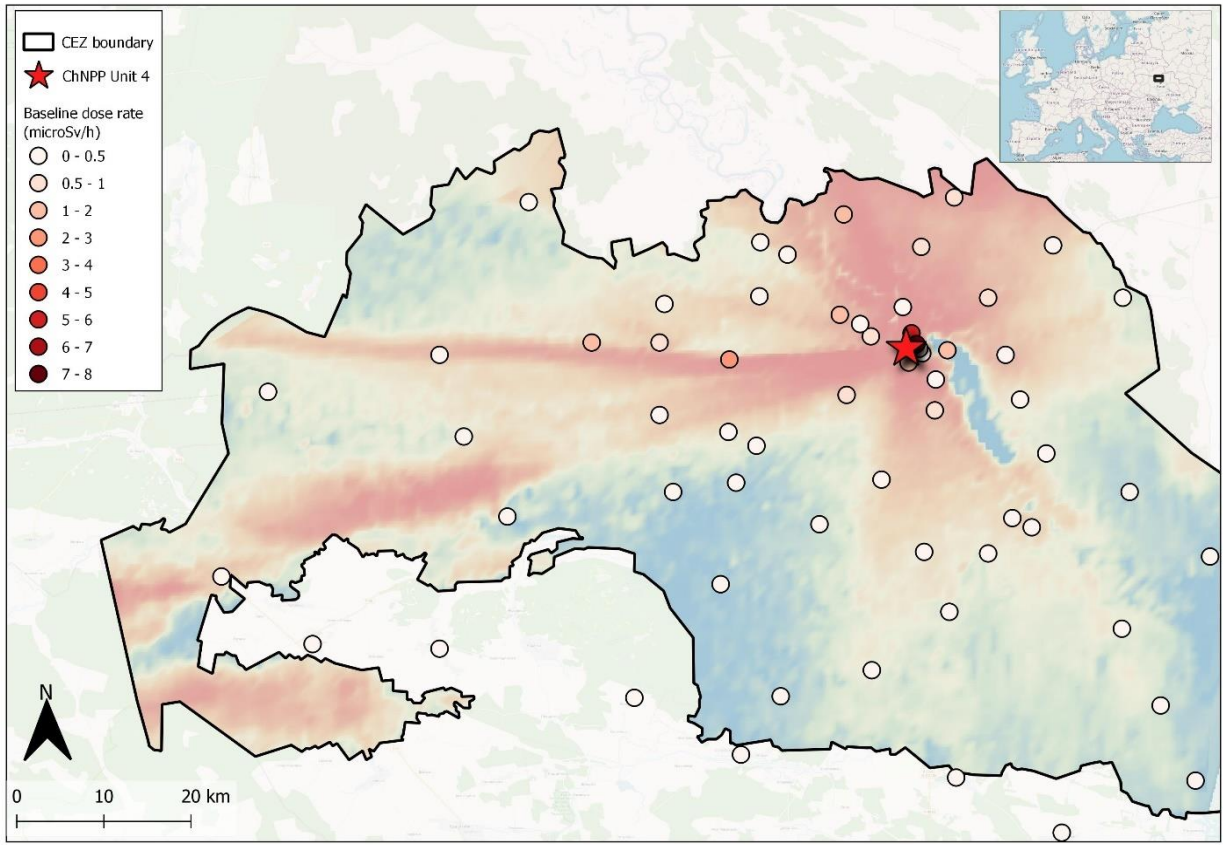

**Figure 1.** Gamma radiation baseline dose rates (µSv h$^{-1}$) reported by the CRMS detector network that monitors the 2600 km$^2$ Ukrainian part of the Chornobyl Exclusion Zone (CEZ). Note that the dose rate categories presented and the associated colour scheme are different to subsequent figures to better present the spatial variability in baseline gamma dose rates across the CEZ. These baseline dose rate data were obtained from saveecobot.com for the week prior to the dose rate increases on 24$^{th}$ and 25$^{th}$ February 2022. Belarus is located directly to the north of the CEZ and Kyiv is approximately 100 km south. The detector towards Slavutych (to the east of the CEZ) are not shown and did not report dose rate increases. The base map colouration within the CEZ boundary shows $^{137}$Cs deposition, decay corrected to present, from the 1986 accident at Unit 4 of the Chornobyl Nuclear Power Plant; light blue is low deposition (tens of kBq m$^{-2}$) and darkest red is highest deposition (thousands of kBq m$^{-2}$). Base map and data from OpenStreetMap and OpenStreetMap Foundation, which is made available under the Open Database License (CC BY-SA 2.0); © OpenStreetMap contributors.

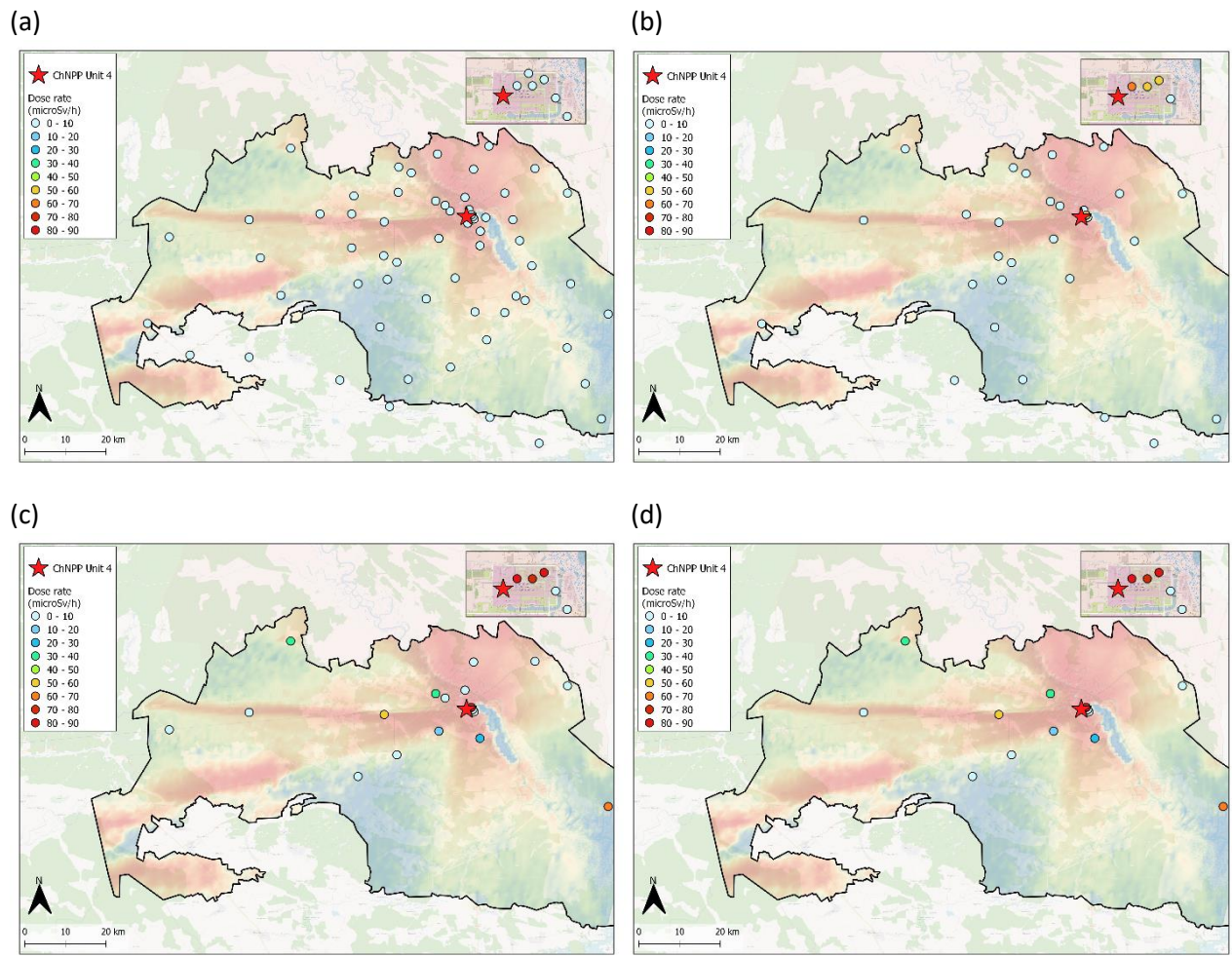

**Figure 2.** Dose rates (μSv h$^{-1}$) reported by CRMS network, shown using the same scale to facilitate comparison for: (a) baseline; (b) increase over baseline on 24$^{th}$ February 2022; (c) increase over baseline on 25$^{th}$ February 2022; and (d) increase above baseline between 09:20 and 10:50 on 25$^{th}$ February 2022. Inset maps show results for detectors closest to the Chornobyl Nuclear Power Plant. Base map and data from OpenStreetMap and OpenStreetMap Foundation, which is made available under the Open Database License (CC BY-SA 2.0); © OpenStreetMap contributors.

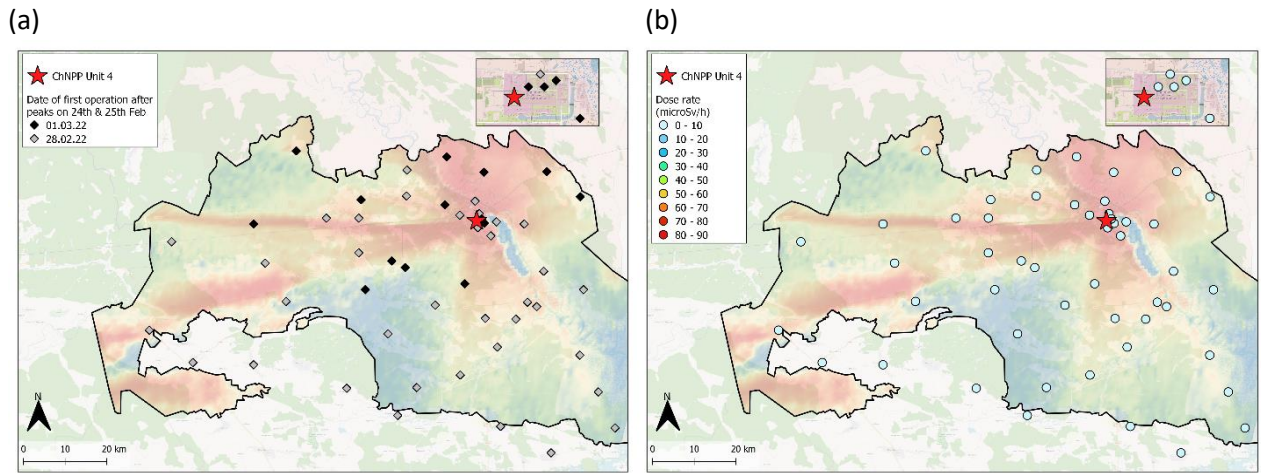

**Figure 3.** CRMS detector network after 25th February 2022, showing: (a) the date each detector first reported dose rate data after the network went offline on 25th February 2022; and (b) gamma dose rates reported (µSv h$^{-1}$), using the same scale categories as Figure 2. Base map and data from OpenStreetMap and OpenStreetMap Foundation, which is made available under the Open Database License (CC BY-SA 2.0); © OpenStreetMap contributors.

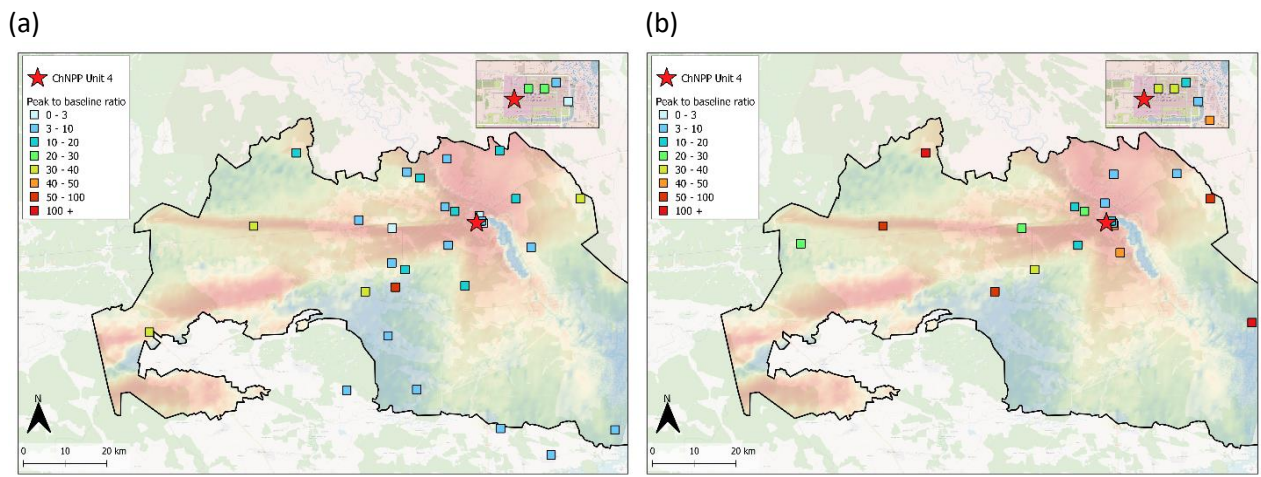

**Figure 4.** Peak to baseline gamma dose rate ratio on: (a) 24th February 2022; and (b) 25th February 2022. Our analysis shows that anything above a ratio of three (0-3 is the first category on the figure scale) cannot be explained by resuspension of contaminated soil due to military vehicle movements. Base map and data from OpenStreetMap and OpenStreetMap Foundation, which is made available under the Open Database License (CC BY-SA 2.0); © OpenStreetMap contributors.

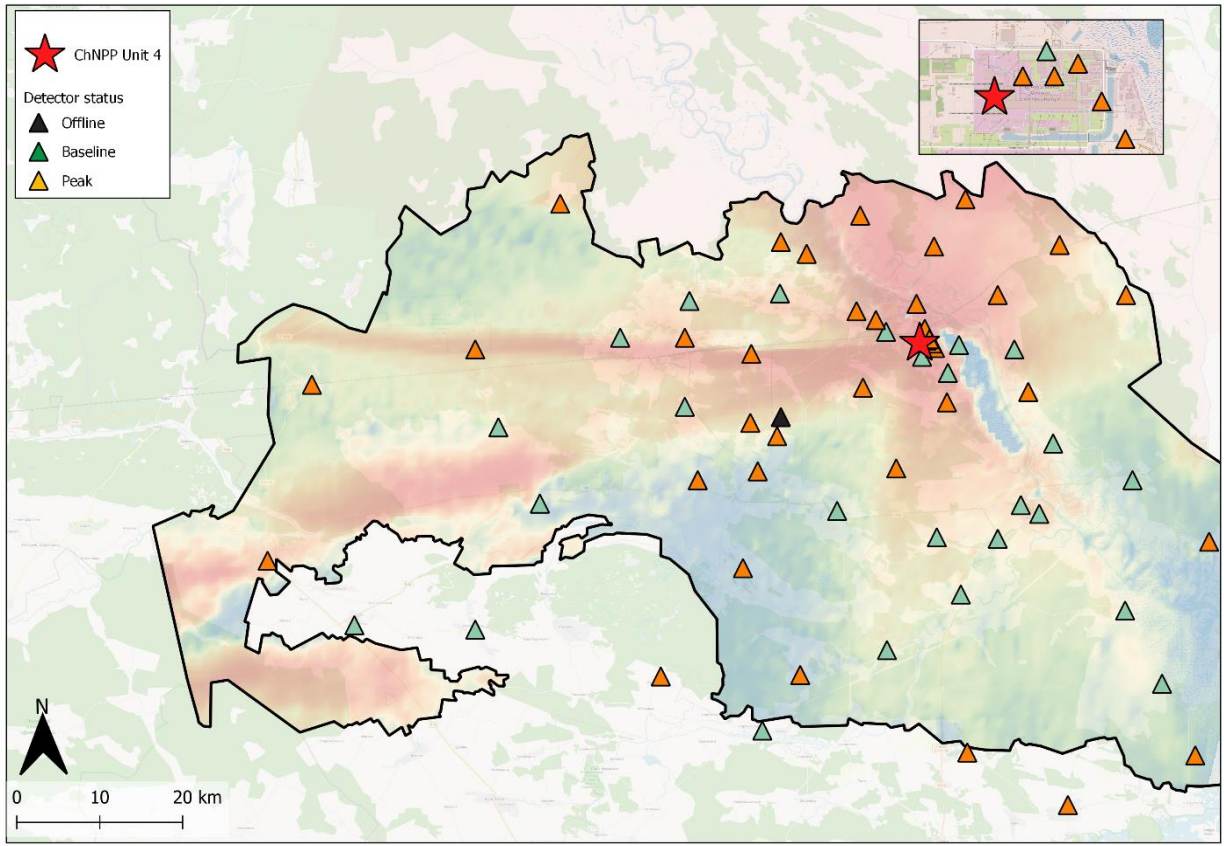

**Figure 5.** Status of detectors within the CRMS network during the period of anomalous readings. One detector towards the centre of the CEZ was offline throughout this period. Detectors that reported increased (peak) dose rates were interspersed with detectors that only reported baseline dose rates. Inset maps show results for detectors closest to the Chornobyl Nuclear Power Plant. Base map and data from OpenStreetMap and OpenStreetMap Foundation, which is made available under the Open Database License (CC BY-SA 2.0); © OpenStreetMap contributors.